\documentclass[letter]{aa} 
\usepackage{graphicx}
\usepackage{txfonts}
\usepackage{longtable}
\usepackage{lscape}
\usepackage{natbib}
\bibpunct{(}{)}{;}{a}{}{,} 
\newcommand\kms{{\rm\,km\,s^{-1}}}

\newcommand\ulsr{U_{\rm LSR}}
\newcommand\vlsr{V_{\rm LSR}}
\newcommand\wlsr{W_{\rm LSR}}

\newcommand\teff{T_{\rm eff}}

\begin{document}

\title{
The first chemical abundance analysis of K giants\\
in the inner Galactic disc\thanks{
This paper includes data gathered with the 6.5 meter Magellan
Telescopes located at the Las Campanas Observatory, Chile
}
}

\author{
T. Bensby\inst{1}
\and
A. Alves-Brito\inst{2}
\and
M.S. Oey\inst{3}
\and
D. Yong\inst{4}
\and
J. Mel\'endez\inst{5}
}

\institute{
European Southern Observatory, Alonso de Cordova 3107, Vitacura, 
Casilla 19001, Santiago 19, Chile
\and
Departamento de Astronom\'ia y Astrof\'isica, Pontificia Universidad 
Cat\'olica de Chile, Santiago, Chile
\and
Department of Astronomy, University of Michigan, Ann Arbor, 
MI 48109-1042, USA
\and
Research School of Astronomy and Astrophysics, Australian National  University,
Weston, ACT 2611, Australia
\and
Centro de Astrof\'{\i}sica, Universidade do Porto, Rua das Estrelas, 
4150-762 Porto, Portugal
}

\date{Received 16 April 2010 / Accepted 11 June 2010}

\offprints{T. Bensby, \email{tbensby@eso.org}}

  \abstract
  %
  {} 
   {
   The elemental abundance structure of the Galactic disc has been
   extensively studied in the solar neighbourhood
   using long-lived stars such as F and G dwarfs or
   K and M giants. These are stars whose atmospheres preserve the 
   chemical composition of their natal gas clouds, and are hence
   excellent tracers of the chemical evolution of the Galaxy. As far as
   we are 
   aware, there are no such studies of the inner Galactic disc,
   which hampers our ability to constrain and trace the origin and 
   evolution of the Milky Way. Therefore, we aim in this study 
   to establish the elemental abundance trend(s) of the disc(s) in 
   the inner regions of the Galaxy.
   }
   {
   Based on equivalent width measurements in high-resolution spectra
   obtained with the MIKE spectrograph on the Magellan II telescope
   on Las Campanas in Chile, we determine elemental abundances
   for 44 K-type red giant stars in the inner Galactic disc, located
   at Galactocentric distances of 4-7\,kpc. The analysis method is 
   identical to the one recently used on red giant stars in the 
   Galactic bulge and in the nearby thin and thick discs, enabling 
   us to perform a 
   truly differential comparison of the different stellar populations.
   }
   {
   We present the first detailed elemental abundance study of
   a significant number of red giant stars in the inner Galactic
   disc.  We find that these inner disc stars show the same type 
   of chemical and kinematical dichotomy as the thin and thick 
   discs show in the solar neighbourhood.  The abundance trends
   of the inner disc agree very well with those of the
   nearby thick disc, and also to those of the Bulge.
   The chemical similarities between the Bulge and the 
   Galactic thick disc stellar populations indicate that they have
   similar chemical histories, and any model trying to 
   understand the formation and evolution of either of the two
   should preferably incorporate both of them.
   }
{}

   \maketitle

\section{Introduction}

The inner Galactic disc is one of the least studied 
regions of the Milky Way because of high interstellar extinction and 
contamination by background Bulge stars. Apart from a few studies 
of bright hot OB stars \citep[e.g.,][]{daflon2004} and Cepheids
\citep[e.g.,][]{luck2006}, which both trace the most recent young disc 
stellar population, almost no information is available about the detailed 
abundance structure of the inner Galactic disc. Open questions
are for instance, whether the inner Galactic disc show the same clear 
kinematic and chemical 
dichotomy as the Galactic disc in the solar neighbourhood, where the 
thin and thick discs stand out as two distinct stellar populations?

Recent studies have revealed that the Galactic bulge and the Galactic 
thick disc have very similar abundance trends which reflect similar, 
and possibly even a shared, chemical histories
\citep{melendez2008,bensby2009,bensby2010,alvesbrito2010}. 
A restriction of these studies is that their thick disc samples have 
been observed in the solar neighbourhood, and if the Bulge has a 
secular origin \citep[e.g.,][]{kormendy2004,howard2009}, models 
show that it likely has to be from gas and stars in 
the inner parts of the Galactic disc \citep[e.g.,][]{rahimi2010}. 
Both the inner and the local disc will help us put constraints on how 
these Galactic components formed, if we can verify the existence
of an inner Galactic thick disc and differentially compare
it with the Bulge.

Here we will present the first results regarding detailed elemental 
abundances in 44 red giant stars that are located at 4-7\,kpc from 
the Galactic centre. They have been analysed using the same 
method as in the recent study of red giants in the Bulge and 
nearby thin and thick discs by \cite{alvesbrito2010}. We will focus 
on four $\alpha$-elements (Mg, Si, Ca and Ti) and omit most of the 
analysis details and results for other iron-peak elements 
for a coming publication.

\section{Sample selection and observations}

\begin{figure}
\resizebox{\hsize}{!}{
\includegraphics[bb=18 155 592 560,clip]{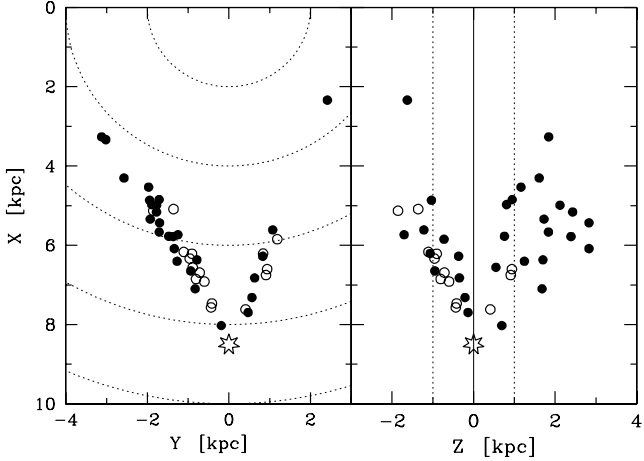}}
\caption{The location of the stars in Galactic $X$, $Y$, and
$Z$ coordinates (distances based on spectroscopic parallaxes).
Symbols as in Fig.~\ref{fig:toomre}.
\label{fig:astrometry} 
}
\end{figure}

One of the caveats in trying to observe the inner Galactic disc in 
the direction of the Galactic centre is that it is very likely that 
the sample will be contaminated by background Bulge stars. However, 
by pointing towards regions on either side of the Bulge, contamination 
is avoided even if the estimated distances are 
greatly in error. Therefore, our targets are located at Galactic 
longitudes $330\degr-340\degr$ and $20\degr-30\degr$ (see left panel 
of Fig.~\ref{fig:astrometry}). 

Because dwarf stars at these distances are too faint to be observed
with high-resolution spectrographs we targeted  bright
red giants. There is a clear separation between dwarfs and giants 
in the de-reddened$^{1}$ $(J-K)$ and $(J-H)$ colour space 
\citep{bessel1988}, and we utilised the selection criteria
of \cite{majewski2003}, who successfully selected distant K and M 
giants from the 2MASS catalogue. To make the sample 
as homogeneous as possible and use early spectral types (to avoid 
TiO bands that gets strong for later types), we selected stars with 
$0.85<(J-K)_{0}<0.88$. This is the intrinsic colour 
for a K4 giant \citep{bessel1988}. The corresponding intrinsic 
$(V-K)$ colour for a K4 giant is 3.26 \citep{bessel1988}, and its 
absolute magnitude is $M_{V}\approx-0.45$ \citep{keenan1999}, giving
$M_{K}=-3.71$. 
The 2MASS $Ks$ magnitudes were transformed to standard $K$ 
magnitudes through $K=Ks+0.044$ \citep{grocholski2002}, and 
after correcting for extinction\footnote{Extinctions were calculated as
($A_{K}$, $E(J-H)$, $E(J-K))=(0.28, 0.34, 0.54) E(B-V)$, where 
$E(B-V)$ is  from the maps by \cite{schlegel1998}, corrected 
using Eq.~1 of \cite{bonifacio2000}.} the distances can be estimated 
using: $K_{0}-M_{K}=-5+5\log d$.

We selected 44 K giants from the 2MASS catalogue that had
estimated Galactocentric distances of 3-7\,kpc. During two observing 
runs in 2007 May and July, high-resolution spectra were obtained 
for all 44 giants with the MIKE spectrograph at the Magellan II 
telescope on Las Campanas in Chile, using a $0.5\arcsec$ wide slit. 
This resulted in spectra with $R\approx 55\,000$, covering the 
entire optical spectrum from 3500 to 10\,000\,{\AA}. Typical 
signal-to-noise ratios are $S/N\approx100$ pixel$^{-1}$ at 6000\,{\AA}.  

\section{Analysis}

Stellar parameters and elemental abundances were determined using
exactly the same spectroscopic methods as outlined in 
\cite{alvesbrito2010}. In short, the analysis is based on 
equivalent width measurements and the ATLAS9
model stellar atmospheres by \cite{castelli1997}. The effective 
temperature ($\teff$)
is found by requiring an excitation balance of the \ion{Fe}{i} line
abundances; surface gravity ($\log g$) by requiring ionisation 
balance between abundances from \ion{Fe}{i} and \ion{Fe}{ii} lines; 
and the microturbulence ($\xi_{\rm t}$) by requiring that the
\ion{Fe}{i} line abundances from are independent of reduced 
line strength.

\begin{figure}
\resizebox{\hsize}{!}{
\includegraphics[bb=18 155 592 545,clip]{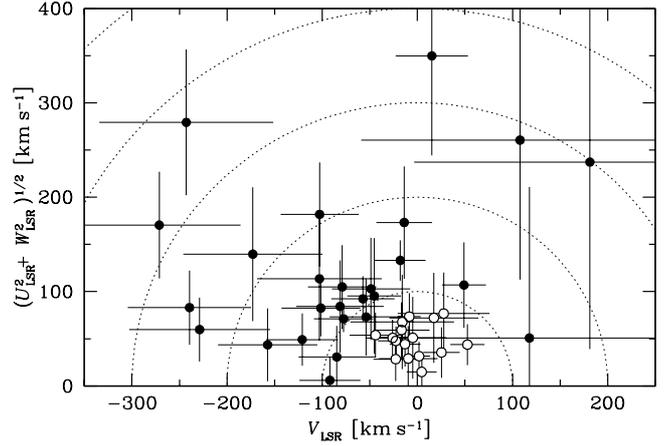}}
\caption{Toomre diagram for 43 of the 44
stars (one does not have measured proper motions). Open circles 
indicate stars that move on 
more or less circular orbits confined to the plane 
($v_{\rm tot}<85\,\kms$), and filled circles those that
move on orbits that are highly eccentric and/or reach far from 
the Galactic plane ($v_{\rm tot}>85\,\kms$).
\label{fig:toomre} 
}
\end{figure}

We find that all stars have effective temperatures in the range 
$4000<\teff <4500$\,K and surface gravities in the range 
$1<\log g < 2.5$, i.e. typical of K giant stars.
Typical  uncertainties are 75\,K in $\teff$, 0.3\,dex in $\log g$,
and 0.2\,$\kms$ in $\xi_{\rm t}$, and  
$\rm \sigma_{[Fe/H]}=0.14$, $\rm \sigma_{[Mg/Fe]}=0.07$, 
$\rm \sigma_{[Si/Fe]}=0.10$, $\rm \sigma_{[Ti/Fe]}=0.13$,
and $\rm \sigma_{[Ca/Fe]}=0.14$ in the 
abundance ratios.

With spectroscopic stellar parameters at 
hand, ``spectroscopic" parallaxes were re-calculated through
\begin{equation}
\log\pi = 0.5([g] - [\mathcal{M}] - 4[T])-0.2(K+BC_{K}-A_{K}+0.25).
\label{eq:gravity}
\end{equation}
Here the notation $[X]\equiv\log(X/X_{\odot})$, and the bolometric 
correction is given by $BC_{K} = -6.75\log(\teff/9500)$ 
\citep{buzzoni2010}. We assume that the giants all have 
$\mathcal{M}=1$\,M$_{\sun}$. Assuming that the uncertainties in 
the stellar parameters and the reddening correction are uncorrelated,
the uncertainties in the distances are calculated to be 30\,\%.
Then we calculated the space velocities 
($\ulsr$, $\vlsr$, and $\wlsr$) using our spectroscopic 
parallaxes, the proper motions from the UCAC3 catalogue 
\citep{zacharias2010}, and radial velocities as measured from the 
spectra. Uncertainties in the space velocities were calculated based
on the assumption that the uncertainties in  the distances and the
proper motions are uncorrelated.
Finally, Galactic orbits were calculated 
with the {\sc grinton} integrator \citep{carraro2002,bedin2006}, 
which gives the minimum and maximum distances from the Galactic centre 
($R_{\rm min}$ and $R_{\rm max}$), the maximum distance from the Galactic 
plane ($Z_{\rm max}$), and the eccentricity of the orbit ($e$).

\section{Results and discussion}

\subsection{Distinct populations in the inner disc?}

In the Toomre diagram in Fig.~\ref{fig:toomre} 
the stars have been coded according to the simple
assumption that those with $v_{\rm tot}>85\,\kms$ 
are thick disc stars, and those with lower velocities  are thin disc 
stars \citep[e.g.,][]{fuhrmann2004}. Because we do not know the properties 
of the inner thick disc, the coding should not be taken literally.
It is also obvious that the errors in the calculated space velocities 
make this classification uncertain.
Hence, we just coded those stars that move on more circular 
orbits and those that have more kinematically hot orbits.
Below we 
will call them kinematically hot stars (black 
circles) and kinematically cold stars (empty circles).

\begin{figure}
\resizebox{\hsize}{!}{
\includegraphics[bb=18 155 592 718,clip]{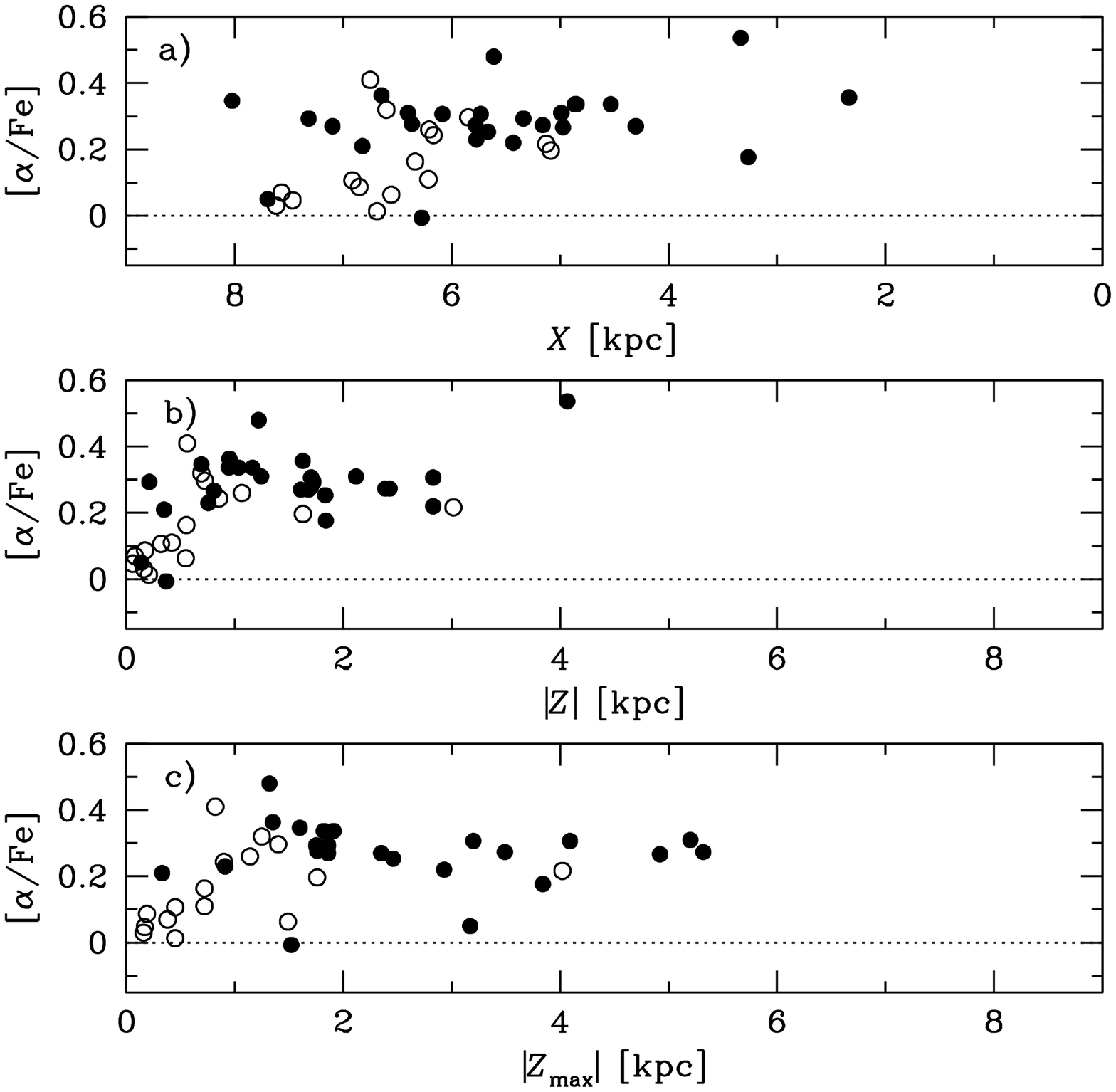}}
\resizebox{\hsize}{!}{
\includegraphics[bb=18 155 592 515,clip]{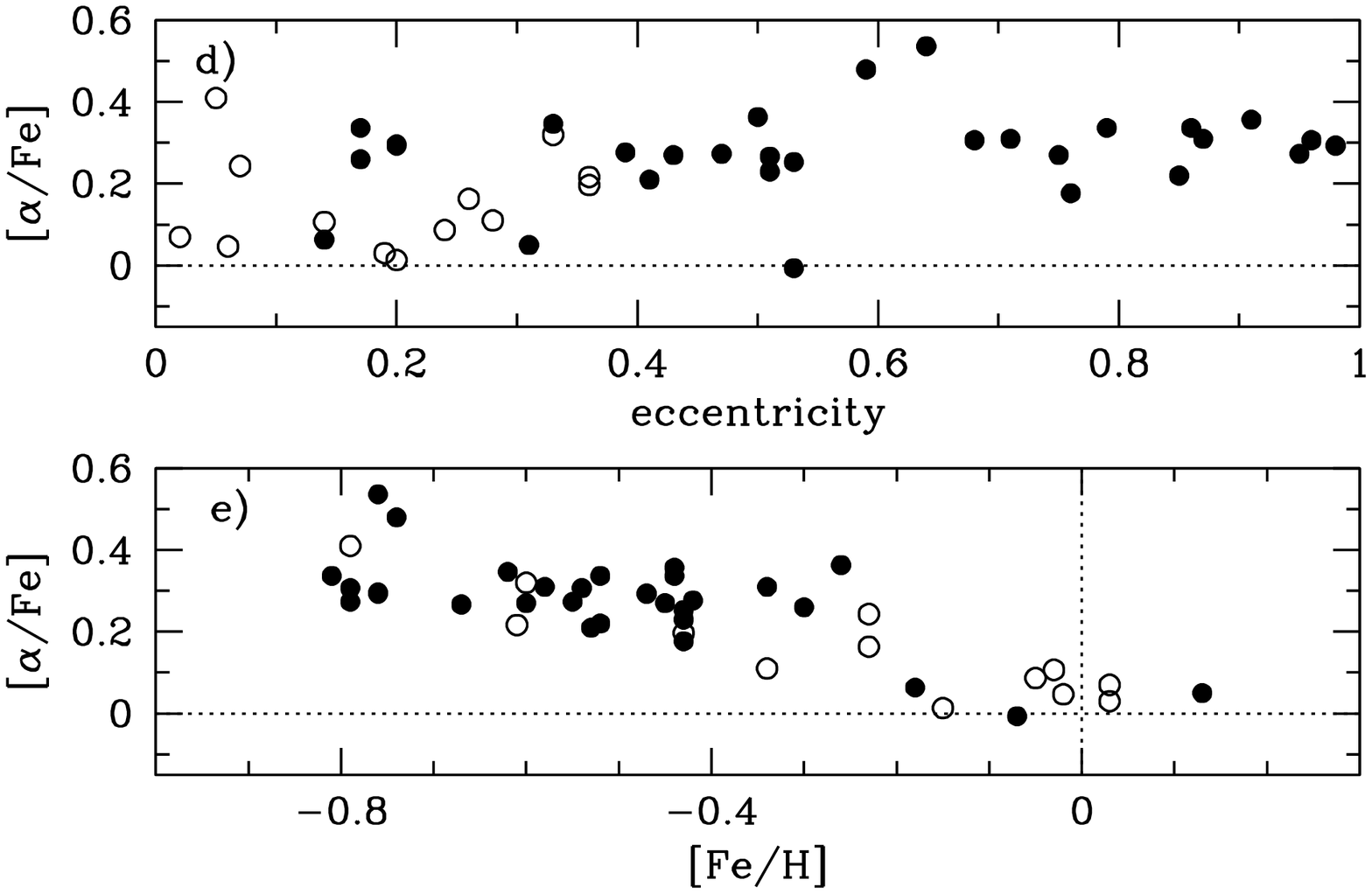}}
\caption{[$\alpha$/Fe] versus $X$, $|Z|$, $Z_{\rm max}$, $e$, and [Fe/H].
Symbols as in Fig.~\ref{fig:toomre}.
\label{fig:alfazmax} 
}
\end{figure}

In Fig.~\ref{fig:alfazmax}a we see that stars with distances greater
than 2.5\,kpc from the Sun have consistently high $\rm [\alpha/Fe]$ 
values ($\rm \alpha\equiv (Mg + Si + Ti)/3$)).
These distant stars are  
all located around or more than 1\,kpc from the plane 
(Fig.~\ref{fig:astrometry}), and they are essentially all
kinematically hot stars. 
Then we see in Fig.~\ref{fig:alfazmax}b that all stars located
more than 1\,kpc from the plane have high $\rm [\alpha/Fe]$
values, and that most of these are kinematically hot stars. 
We also see a few kinematically hot stars
that are located close to the
plane and also have low $\rm [\alpha/Fe]$ values. However,
Fig.~\ref{fig:alfazmax}c shows that these stars have kinematic
properties that allow them to reach as far as $\sim$2\,kpc from 
the plane. At the same time, stars that are close to the
plane, have low $\rm [\alpha/Fe]$ values, and are 
kinematically cold, stay within 1\,kpc from the plane.
We also note that we have a few stars with high $\rm [\alpha/Fe]$
values, which are kinematically hot, but which remain close
to the the plane. These are stars that have highly
eccentric orbits.
Figure~\ref{fig:alfazmax}d then shows that stars that move on 
highly eccentric orbits all have high $\rm [\alpha/Fe]$ values,
and are all classified as kinematically hot. 
For stars with less eccentric orbits 
there is a gradual decrease in $\rm [\alpha/Fe]$ as the orbits become 
more circular. With a few exceptions, the stars with the least
eccentric orbits have the lowest $\rm [\alpha/Fe]$ values. 
Figure~\ref{fig:alfazmax}e  shows that stars with low [Fe/H] have high
$\rm [\alpha/Fe]$, with a flat trend that eventually
starts to decrease for metallicities higher than 
$\rm [Fe/H]\approx-0.3$.  Also, the stars with cold kinematics 
generally have higher [Fe/H] and lower $\rm [\alpha/Fe]$.

These connections and correlations between kinematics and chemistry 
that we see for the inner disc sample is what we see for disc 
stars in the solar neighbourhood. Stars
with orbits that are highly eccentric and/or reach far from the
plane generally have high $\rm [\alpha/Fe]$ values, and those
on more circular orbits, which stay closer to the plane, have
low $\rm [\alpha/Fe]$ values. Stars with these properties
are generally classified as thick disc and thin disc stars,
respectively \citep[e.g.,][]{bensby2005,fuhrmann2004}. 
That we see the same correlations in the inner Galactic disc
strongly suggests that we have two distinct disc populations 
also in the inner disc, an inner thin disc and an inner thick disc,
similar to those in the solar neighbourhood.

\subsection{The Galactic bulge -- thick disc connection}

\begin{figure*}
\resizebox{\hsize}{!}{
\includegraphics[bb=10 204 570 420,clip]{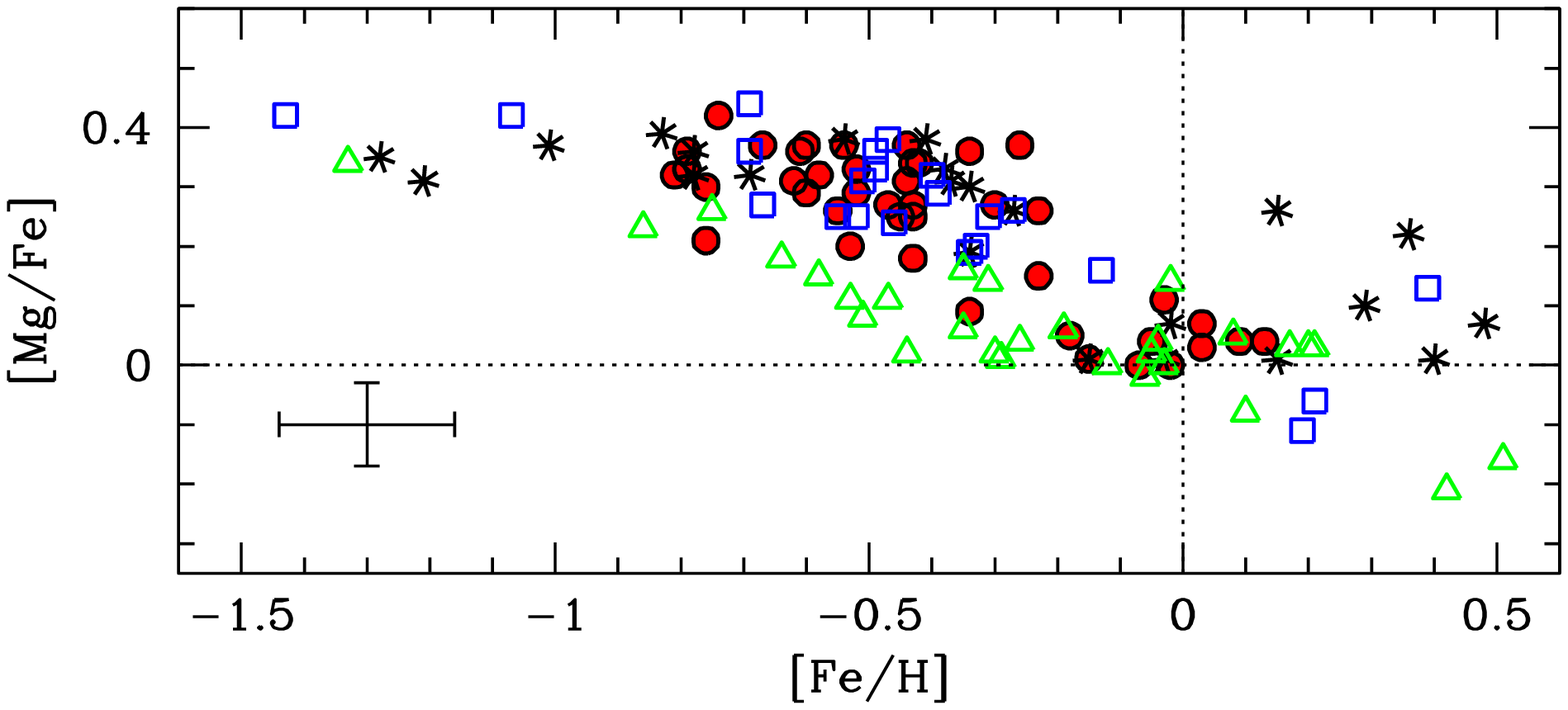}
\includegraphics[bb=18 204 592 420,clip]{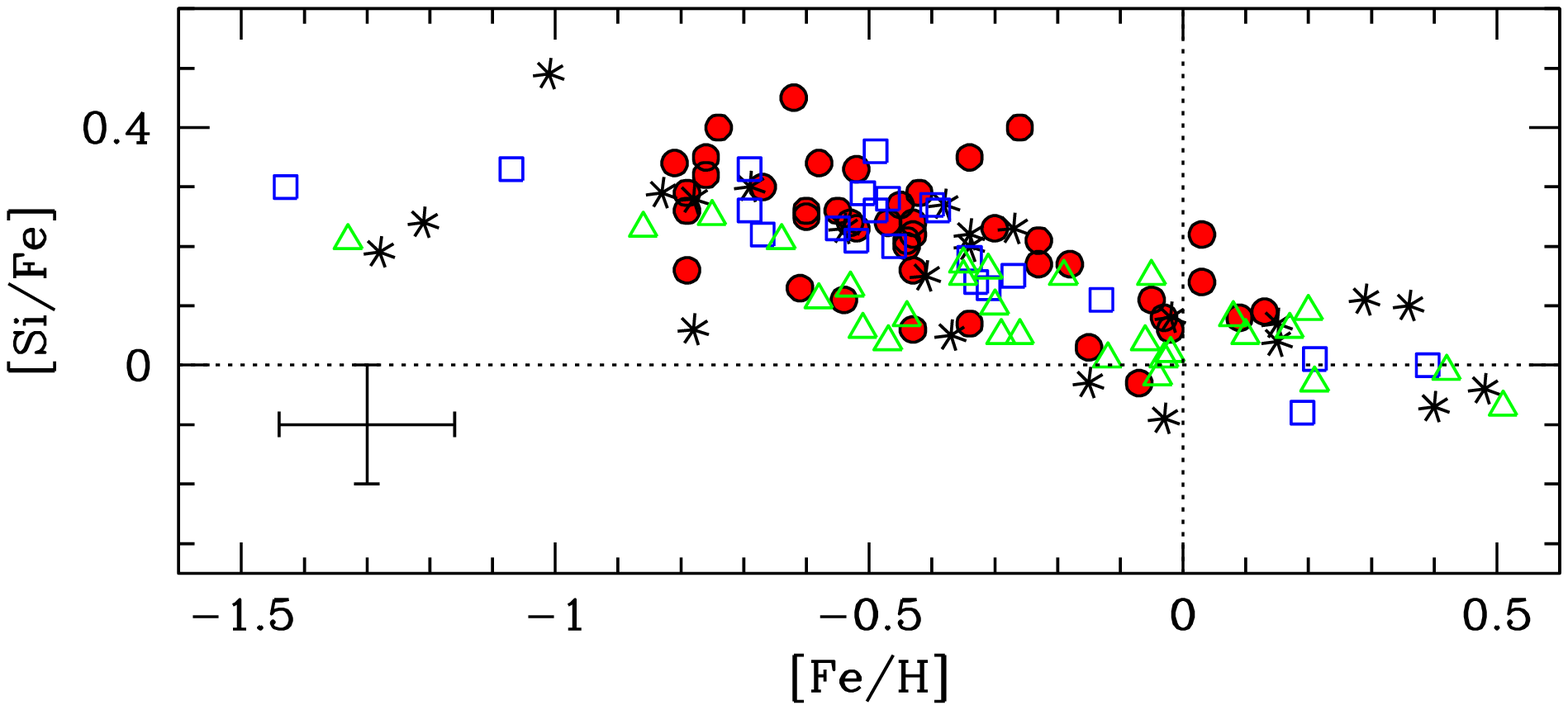}}
\resizebox{\hsize}{!}{
\includegraphics[bb=10 155 570 405,clip]{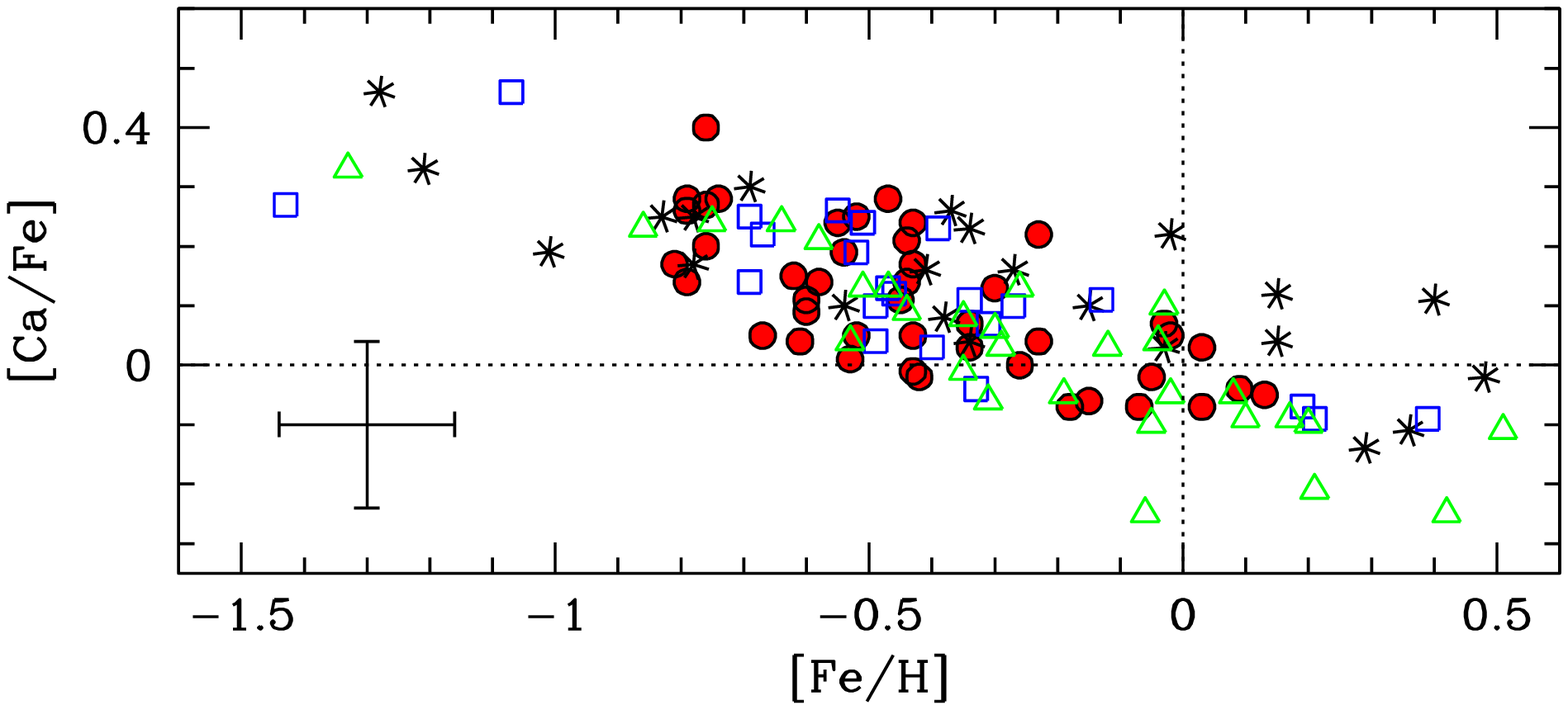}
\includegraphics[bb=18 155 592 405,clip]{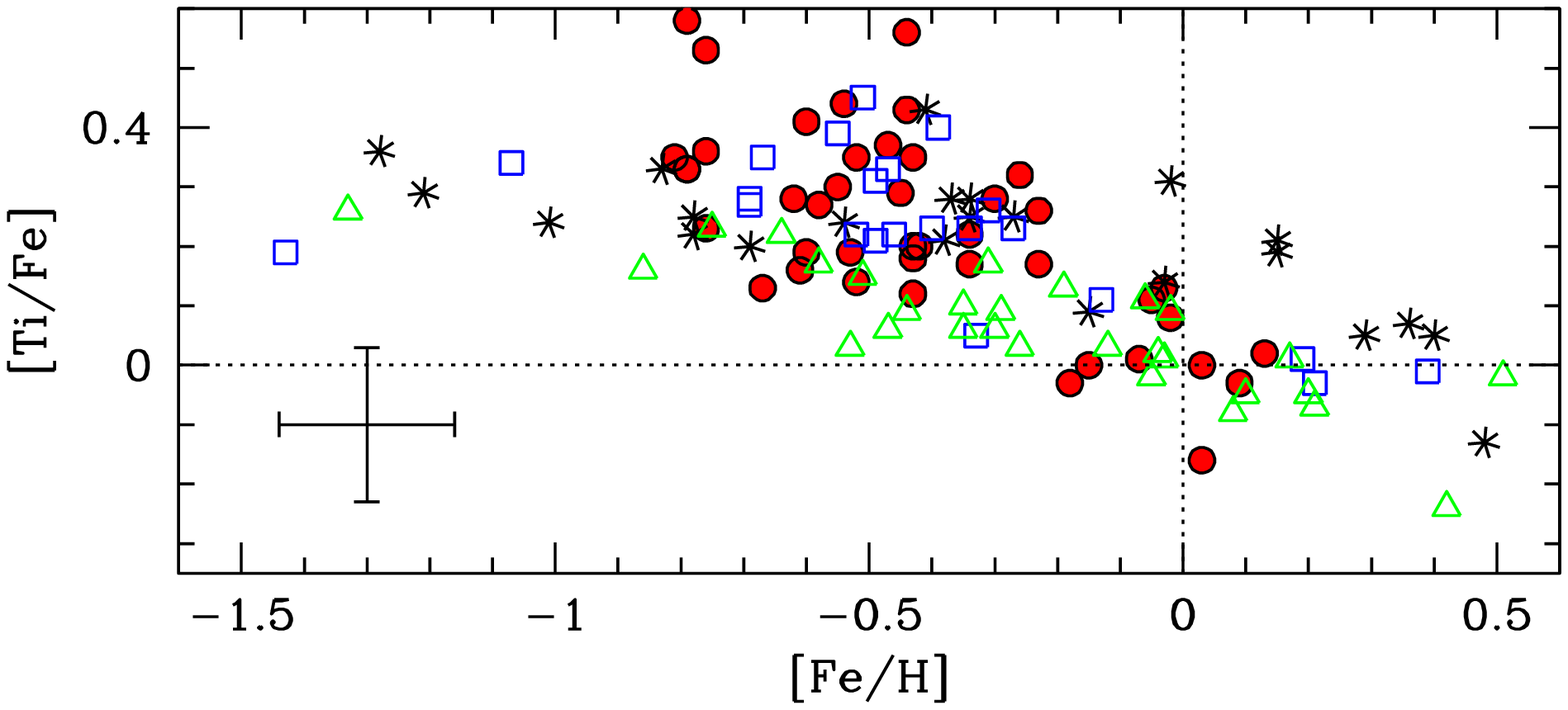}}
\caption{Abundance trends for our inner disc giants
(red filled circles) together with Bulge giants 
(asterisks), nearby thin disc giants (green empty triangles);
and nearby thick disc giants (blue empty squares), all
from \cite{alvesbrito2010}. Typical error bars are shown in each plot.
\label{fig:haltplottar} 
}
\end{figure*}

In Fig.~\ref{fig:haltplottar} we show the detailed abundance trends
for four $\alpha$-elements, comparing the 44 inner disc K giants 
to the Bulge giants and nearby thin and thick disc giants from 
\cite{alvesbrito2010}. We emphasise that all stars in these plots 
have been analysed with the exact same methods, allowing truly 
differential comparisons between the different populations.

We note that especially the Mg abundance trend
shows very little scatter, and that the inner disc giants have 
high [Mg/Fe] ratios for $\rm [Fe/H]<-0.3$ and lower enhancements 
for higher [Fe/H]. This is a 
signature of enrichment by massive stars at low metallicities, and 
a delayed contribution from low mass stars at higher metallicities, 
consistent with the same signature seen in the nearby thick disc 
\citep[e.g.,][]{feltzing2003}. This points to the existence of an 
inner thick disc and moreover that this thick disc does not differ much 
in terms of abundance trends, from the thick disc we see in the solar
neighbourhhod. The same trend that is seen for Mg can also be seen in 
the Si and Ti plots, but with larger scatters. No clear trend can be 
seen in the Ca plot. 

Furthermore, the abundance trends of the inner disc appear to be
very similar to those of the Bulge. This inevitably points to 
a possible connection between the
thick disc and the Bulge, implying they both might have formed at the
same time \cite[e.g.,][]{genzel2008}, sharing a similar star-formation 
rate and initial mass function. A possible scenario could be that 
the sub-solar part of the Bulge has a secular origin, and has formed 
from inner disc material \citep[e.g.,][]{shen2010}.  

The agreement between the Bulge and the thick disc has recently also 
been seen in studies that compare Bulge stars with nearby thick disc 
stars. For instance, \cite{bensby2010} presented a detailed abundance
analysis of 15 microlensed dwarf stars in the Galactic bulge.
These stars were found to share the same abundance trends as 
were traced by 
kinematically selected thick disc dwarf stars in the solar neighbourhood 
\citep[][and 2010, in prep.]{bensby2003,bensby2005}. Similarly, 
\cite{melendez2008}  and \cite{alvesbrito2010} found very good 
agreement between the abundance trends of red giants in the Bulge and 
thick disc red giants in the solar neighbourhood (see also 
\citealt{ryde2010}). Similar to this study, the analysis methods of 
these studies are internally fully consistent (same methods, model 
stellar atmospheres, atomic data, etc.). They compare dwarfs 
with dwarfs, and giants with giants. Other studies of red giants in 
the Bulge \citep[e.g.][]{fulbright2007,zoccali2006,lecureur2007}
have found that the Bulge is significantly more $\alpha$-enhanced
at higher metallicities than thin and thick disc stars. As discussed
in \cite{bensby2010} and \cite{alvesbrito2010}, it is likely that 
those studies suffer from problems with the analysis (especially
line blending). They also compare their Bulge giant samples with
disc dwarf samples. The combined effect is that their Bulge stars 
seem spuriously more 
enhanced in the $\alpha$-elements than the thick disc stars.

We further note that none of the inner disc giants are as metal-rich 
as some of the most metal-rich Bulge giants. 
As the metallicity distribution of the thick disc peaks at
$\rm [Fe/H]\approx -0.6$ \citep{carollo2010}, it is not surprising 
that our sample does not contain many metal-rich (thick disc) stars.
Instead the upper 
metallicity limit appears to be close to, or slightly above, solar 
values (similar to what is seen for nearby thick disc dwarfs stars, 
\citealt{bensby2007letter2}).
A possible connection between the thick disc and the
metal-rich part of the Bulge is therefore dubious.
In that case, the metal-rich parts of the Bulge must have 
another origin, which possibly could be from accreted 
(extra-galactic?) material 
\citep[see, e.g., the models by][]{rahimi2010}.
Evidence for two co-existing formation scenarios within the Bulge
was recently shown by \cite{bensby2010} and 
\cite{babusiaux2010}. With our result for the inner Galactic disc,
the bonds between the the metal-poor part of the Bulge and the Galactic 
thick disc have grown even stronger.

\section{Summary}

We have presented the first detailed elemental abundance study of K 
giants in the inner Galactic disc. Our sample consists of 44 stars 
positioned 4-7\,kpc from the Galactic centre, and up to 3\,kpc from 
the Galactic plane. The three main results are: 
\begin{itemize}
\item based on elemental abundances and kinematics,
we find it likely that the inner Galactic disc has two distinct
stellar populations: a thin disc and a thick disc;
\item the abundance trends of the inner Galactic thick disc 
are similar to those of the thick disc in the solar neighbourhood;
\item we confirm, now using inner disc giants, the chemical
similarity between the Galactic thick disc and the 
metal-poor Bulge.
\end{itemize}
Finally, our results do not preclude the possibility that the local thick disc could be in part produced by radial mixing of inner disc stars \citep{schonrich2009b}.

In a forthcoming paper we will present the analysis
of the current sample in detail and also add abundance results for more
elements. That study will also include another similar
sample of giant stars, but located in the outer Galactic disc.

\begin{acknowledgement}
TB and MSO acknowledge support by the National Science 
Foundation, grant AST-0448900. AAB acknowledges grants from 
FONDECYT (process 3100013). JM is supported by a Ci\^encia 2007 
contract (FCT/MCTES/Portugal and POPH/FSE/EC) and acknowledges 
support from PTDC/CTE-AST/65971/2006 (FCT/Portugal).
\end{acknowledgement}

\bibliographystyle{aa}
\bibliography{referenser}

\end{document}